\documentclass[12pt,reqno,a4paper,pagebackref]{amsart}
\usepackage{graphicx}
\usepackage{fancyhdr}
\usepackage{enumerate}
\usepackage{amsmath}

\usepackage[multiple]{footmisc}

\usepackage{amsthm}
\usepackage{mathabx}
\usepackage{amssymb}
\usepackage{relsize}
\usepackage{pdfsync}
\usepackage[colorlinks=true,linkcolor=blue,citecolor=magenta]{hyperref}

\usepackage[normalem]{ulem}

\usepackage{xcolor}

\usepackage{cancel}

\usepackage{calligra}

\usepackage{tikz}

\usepackage{changebar}
\usepackage{pdfsync}
\usetikzlibrary{decorations.pathmorphing,patterns}
\usetikzlibrary{calc,arrows}

\usepackage{pgf}

\usepackage{setspace}

\usetikzlibrary{automata}
\usetikzlibrary{positioning}

\tikzset{
    state/.style={
           rectangle,
           rounded corners,
           draw=black, very thick,
           minimum height=2em,
           inner sep=2pt,
           text centered,
           },
}

\usepackage{MnSymbol}

\usepackage{changebar}
\usepackage{pdfsync}

\usepackage[normalem]{ulem}

\usepackage{mathtools}

\makeindex
\newcommand{\cal}{\mathcal}
\newcommand{\mc}[1]{{\mathcal #1}}

\newcommand\bx{{\mathbf x}}
\newcommand\bbR{{\mathbb R}}
\newcommand\bbZ{{\mathbb Z}}

\newcommand\lang{\langle\langle}
\newcommand\rang{\rangle\rangle}

\newcommand\E{{\mathbb E}}
\newcommand\bbE{{\mathbb E}}

\newcommand\R{{\mathbb R}}

\newcommand\TT{{\mathbb T}}

\newcommand \ga{\gamma}
\newcommand \om{\omega}

\newcommand\Om{\Omega}

\renewcommand{\ge}{\geqslant}
\renewcommand{\le}{\leqslant}
\newcommand{\dd  }{\mathrm{d}}

\renewcommand{\E}{\mathcal{E}}
\renewcommand{\hat}{\widehat}
\renewcommand{\tilde}{\widetilde}
\renewcommand{\bar}{\overline}
\numberwithin{equation}{section}

\newcommand{\rv}{{\bf r}}
\newcommand{\pv}{{\bf p}}
\newcommand{\qv}{{\bf q}}

\newcommand{\cF}{{\mathcal F}} 

\definecolor{armygreen}{rgb}{0.29, 0.33, 0.13}
\definecolor{britishracinggreen}{rgb}{0.0, 0.26, 0.15}
\definecolor{cadmiumgreen}{rgb}{0.0, 0.42, 0.24}
\definecolor{crimson}{rgb}{0.86, 0.08, 0.24}
\definecolor{darkblue}{rgb}{0.0, 0.0, 0.55}
\definecolor{electricpurple}{rgb}{0.75, 0.0, 1.0}
\definecolor{electriclime}{rgb}{0.8, 1.0, 0.0}
\definecolor{fawn}{rgb}{0.9, 0.67, 0.44}
\definecolor{green(ryb)}{rgb}{0.4, 0.69, 0.2}

\begin{document}

\title[Heat equation with boundaries]{On the Conversion of Work into Heat:
Microscopic Models and Macroscopic Equations}

 \author{Tomasz Komorowski}
 \address{Tomasz Komorowski\\Institute of Mathematics,
   Polish Academy
  Of Sciences\\Warsaw, Poland.} 
\email{{\tt tkomorowski@impan.pl}}

\author{Joel L. Lebowitz}
\address{Joel Lebowitz, Departments of Mathematics and Physics,  Rutgers University}
\email{lebowitz@math.rutgers.edu}
 
 \author{Stefano Olla}
 \address{Stefano Olla, CEREMADE,
   Universit\'e Paris Dauphine - PSL Research University \\
\emph{and}  Institut Universitaire de France\\
\emph{and} GSSI, L'Aquila}
  \email{olla@ceremade.dauphine.fr}

  \author{Marielle Simon}
  \address{Université Claude Bernard Lyon 1, CNRS UMR 5208, Institut Camille Jordan, F-69622 Villeurbanne, France\\
    \emph{and} GSSI, L'Aquila}
  \email{msimon@math.univ-lyon1.fr}

\date{\today {\bf File: {\jobname}.tex.}}

  \begin{abstract}
We summarize and extend some of the results obtained recently for the
  microscopic and macroscopic behavior of
  a pinned harmonic chain, with random velocity flips at Poissonian
  times, acted on by a periodic force {at one end} and in contact
  with a heat bath {at the other end}. Here we consider the case
  where the system is in contact with two
  heat baths at different temperatures and a periodic force is applied at any position.
  This leads in the hydrodynamic
  limit to a heat equation for the temperature profile
  with a discontinuous slope at the position where the
  force acts. Higher dimensional systems, unpinned cases and anharmonic
  interactions are also considered.
\\
  \emph{Dedicated to Errico Presutti for his 80th birthday!}
\end{abstract}



\thanks{{\bf Acknowledgements:} The work of J.L.L. was supported in
  part by the A.F.O.S.R. He thanks the Institute for Advanced Studies
  for its hospitality.
  T.K. acknowledges the support of the NCN grant 2020/37/B/ST1/00426.
  This project is partially supported by the ANR grant MICMOV (ANR-19-CE40-0012)
  of the French National Research Agency (ANR),
  and by the European Union with the program Fonds européen de développement régional.
} 

\maketitle

\section{Introduction}\label{intro}

{Nature has a hierarchical structure with macroscopic behavior
  arising from the dynamics of atoms and molecules. The connection
  between different levels of the hierarchy is however not always
  straightforward, as seen in the emergent phenomena, such as phase
  transitions and heat convection. Establishing in a mathematical
  precise way the connection {between} the different levels is the
  central problem of rigorous statistical mechanics.}

The derivation of macroscopic behavior from microscopic models
by suitable scaling of space and time is a field
of science to which Errico has made seminal contributions
both for equilibrium and nonequilibrium systems.
In this work, which owes a lot to what we have learned from him,
we study the transition from microscopic to macroscopic systems
in the context of the conversion of {work} to heat.
The conversion of mechanical energy into heat was demostrated by Joule's
famous experiment in the 1840's.
Joule dropped weights turning a paddle wheel immersed in water.
The friction generated heat which he could measure and quantify.

In some recent works on this subject we carried out
a rigorous mathematical analysis
of a simple microscopic
model for this common phenomena.
{ In articles \cite{klo22,klo22-2}},
we considered a pinned harmonic chain on which work is done at the right end
by an external periodic force.
This work is converted into heat via an energy current flowing into a heat reservoir at
the left end of the chain.
In order to make this system {mirror} realistic physical systems
with a finite heat conductivity
we added to the bulk dynamics a random reversal
of the velocity of each particle at a rate $\gamma$ (the pure
harmonic crystal is well known to have an infinite heat conductivity, {see
  e.g.~\cite{RLL67}}). {The precise description of the model is
  given in Section \ref{sec:description}.}
Starting with an initial distribution on the phase space, we have shown that the
system approaches a unique periodic state at long times.
We  have also obtained, in the hydrodynamical { diffusive}
scaling limit,
a heat equation for the temperature profile
of the chain.

As a consequence of the presence of the periodic forcing, a constant energy flux,
{equal to the work done by the force},
emerges on the macroscopic scale, as well as
a boundary condition on the derivative of the temperature profile
(Neumann boundary condition), such that Fourier law is satisfied with respect to
this energy flux. In \ref{sec:pinned-dynamics} we review these
results and we present a generalization where the periodic force is
applied at any position inside the system, generating  Neumann type of boundary
conditions in the bulk.
{We should mention the pioneering work by Errico and collaborators
  \cite{dpt11} where \emph{current reservoirs} are attached to the boundary of
  an open symmetric simple exclusion dynamics, that originate non-linear Dirichlet boundary conditions.}

In Section \ref{sec:unpinn-dynam-omeg} we review a work in preparation
about the unpinned dynamics \cite{klos22}. In this situation there are two
locally conserved quantities, energy and volume stretch, and the macroscopic
evolution is governed by a  coupled system of two diffusive equations, see (\ref{eq:linear}) below.
In the absence of periodic forcing this problem was previously studied in \cite{kos3}.
In Section \ref{sec:d-dimension} we review the generalization to higher dimension,
also a work in preparation, as the proof is not a straightforward generalization of the
one-dimensional case. In Section \ref{sec:anharmonic-chains} we describe few results
and some conjectures
about the anharmonic case.

\bigskip
\noindent{\bf Acknowledgements.}  The work of J.L.L. was supported in
  part by the A.F.O.S.R. He thanks the Institute for Advanced Studies
  for its hospitality.
  T.K. acknowledges the support of the NCN grant 2020/37/B/ST1/00426.  This project is partially supported by the ANR grant MICMOV (ANR-19-CE40-0012)
of the French National Research Agency (ANR), and by the European Union with the program Fonds européen de développement régional.

\section{Description of the model}
\label{sec:description}

We consider a pinned chain of $n+1$ harmonic oscillators in contact with
a Langevin heat bath at temperature $T_-$ on the left and another
Langevin heat bath at temperature $T_+$ on the right.
In addition there is a periodic force acting on
the particle labeled by $[n\bar u]$, where $\bar u\in [0,1]$ and $[a]$ denotes the integer part of a positive real number $a$.
The configuration of particle positions and momenta is described by
\begin{equation}
  \label{eq:1}
  (\mathbf q, \mathbf p) =
  (q_0, \dots, q_n, p_0, \dots, p_n) \in \R^{n+1}\times\R^{n+1}. 
\end{equation}
The total energy of the chain is given by the Hamiltonian:
$\mathcal{H}_n (\mathbf q, \mathbf p):=
\sum_{x=0}^n {\cal E}_x (\mathbf q, \mathbf p),$
where the energy of particle $x$ is defined by
\begin{equation}
\label{Ex}
{\cal E}_x (\mathbf q, \mathbf p):=  \frac{p_x^2}2 +
\frac12 (q_{x}-q_{x-1})^2 +\frac{\om_0^2 q_x^2}{2} ,\
\quad x = 0, \dots, n,
\end{equation}
where $\om_0>0$ is the pinning strength.
We adopt the convention that $q_{-1}:=q_0$.

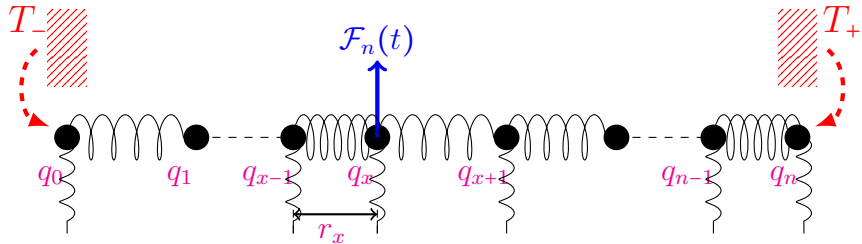
\begin{figure}[h!]

\begin{center}
\begin{tikzpicture}[scale = 0.85]
\node[circle,fill=black,inner sep=1.2mm] (e) at (0,0) {};
\node[circle,fill=black,inner sep=1.2mm] (f) at (2,0) {};
\node[circle,fill=black,inner sep=1.2mm] (g) at (3.5,0) {};
\node[circle,fill=black,inner sep=1.2mm] (h) at (4.8,0) {};
\node[circle,fill=black,inner sep=1.2mm] (i) at (6.8,0) {};
\node[circle,fill=black,inner sep=1.2mm] (j) at (8.5,0) {};
\node[circle,fill=black,inner sep=1.2mm] (k) at (10,0) {};
\node[circle,fill=black,inner sep=1.2mm] (l) at (11.3,0) {};

\draw[dashed] (2,0) -- (3.5,0);
\draw[dashed] (8.5,0) -- (10,0);
\draw[ultra thick, blue, ->] (4.8,0) -- (4.8,1.2);

\draw[thick, <->] (3.5,-1.2) -- (4.8,-1.2);

\draw (-0.25, -0.6) node[] {\color{magenta} $q_{0}$};
\draw (1.75, -0.6) node[] {\color{magenta} $q_{1}$};
\draw (11.1, -0.6) node[] {\color{magenta} $q_{n}$};
\draw (9.61, -0.6) node[] {\color{magenta} $q_{n-1}$};
\draw (3.12, -0.6) node[] {\color{magenta} $q_{x-1}$};
\draw (4.55, -0.6) node[] {\color{magenta} $q_{x}$};
\draw (6.45, -0.6) node[] {\color{magenta} $q_{x+1}$};
\draw[dashed] (3.5,-1.5) -- (3.5,-1);
\draw (4.1, -1.5) node[] {\color{magenta} $r_x$};
\draw[dashed] (4.8,-1.5) -- (4.8,-1);

\draw (-0.6,1.8) node[] {\large\color{red}$T_-$};
\draw (12,1.8) node[] {\large\color{red}$T_+$};
\draw (4.8,1.5) node[] {\color{blue}${\mathcal F}_n(t)$};

\draw[decoration={aspect=0.3, segment length=3mm, amplitude=3mm,coil},decorate] (0,0) -- (2,0);
\draw[decoration={aspect=0.3, segment length=3mm, amplitude=1mm,coil},decorate] (0,0) -- (0,-1.5);
\draw[decoration={aspect=0.3, segment length=1.8mm, amplitude=3mm,coil},decorate] (3.5,0) -- (4.9,0);
\draw[decoration={aspect=0.3, segment length=3mm, amplitude=1mm,coil},decorate] (3.5,0) -- (3.5,-1.5);
\draw[decoration={aspect=0.3, segment length=3mm, amplitude=3mm,coil},decorate] (4.8,0) -- (6.9,0);
\draw[decoration={aspect=0.3, segment length=3mm, amplitude=1mm,coil},decorate] (4.8,0) -- (4.8,-1.5);
\draw[decoration={aspect=0.3, segment length=2.5mm, amplitude=3mm,coil},decorate] (6.8,0) -- (8.6,0);
\draw[decoration={aspect=0.3, segment length=3mm, amplitude=1mm,coil},decorate] (6.8,0) -- (6.8,-1.5);
\draw[decoration={aspect=0.3, segment length=1.8mm, amplitude=3mm,coil},decorate] (10,0) -- (11.4,0);
\draw[decoration={aspect=0.3, segment length=3mm, amplitude=1mm,coil},decorate] (10,0) -- (10,-1.5);
\draw[decoration={aspect=0.3, segment length=3mm, amplitude=1mm,coil},decorate] (11.4,0) -- (11.4,-1.5);

\fill [pattern = north east lines, pattern color=red] (-0.3,0.8) rectangle (0.3,2);
\fill [pattern = north east lines, pattern color=red] (11,0.8) rectangle (11.6,2);
\node (c) at (-0.3,1.5) {};
\node (d) at (-0.1,0.1) {};
 \node (a) at (11.6,1.5) {};
 \node (b) at (11.4,0.1) {};

\draw (c) edge[dashed, ultra thick, red, ->, >=latex, bend right=60] (d);
\draw (a) edge[dashed, ultra thick, red, ->, >=latex, bend left=60] (b);

\end{tikzpicture}
\end{center}
\end{figure}

The  microscopic dynamics of
the process $\{(\mathbf q(t), \mathbf p(t))\}_{t\ge0}$
describing the total chain is  given in the bulk by
\begin{equation} 
\label{eq:flip}
\begin{aligned}
  \dot   q_x(t) &= p_x(t) ,
  \qquad  x\in \{0, \dots, n\},\\
  \dd   p_x(t) &=  \left(\Delta_N q_x-\om_0^2 q_x\right) \dd t-   2
  p_x(t-) \dd N_x(\gamma t) + \delta_{x,[n\bar u]}\cF_n(t)
                 \dd t  ,
\end{aligned} \end{equation} for $x\in \{1, \dots, n-1\}$,  
and at the boundaries the equations are 
\begin{align}
     \dd   p_0(t) &=   \; \Big(q_1(t)-q_0(t) - \om_0^2 q_0(t) \Big) \dd   t
                     -
                    2  \gamma_- p_0(t) \dd t
                    +\sqrt{4  \gamma_- T_-} \dd \tilde w_-(t)
                    \vphantom{\Big(} \label{eq:pbdf} \\
  \dd   p_n(t) &=  \; \Big(q_{n-1}(t) -q_n(t) -\om_0^2 q_n(t) \Big)  \dd   t
                 -  2  \gamma_+ p_n(t) \dd t +\sqrt{4  \gamma_+ T_+}\dd \tilde w_+(t).
                     \vphantom{\Big(}  
\notag
\end{align}
Here {$\Delta_Nq_x=q_{x+1}+q_{x-1}-2q_x$} is the Neumann discrete
Laplacian, corresponding to the choice of {the boundary conditions}
$q_{n+1}:=q_n$ and {$q_{-1}:= q_0$}. Processes $\{N_x(t), x=1,\ldots,n-1\}$
are independent Poisson processes of intensity $1$, while
$\tilde w_\pm(t)$ are two independent standard Wiener processes,
independent of the Poisson processes.
Parameters $\gamma> 0, \gamma_\pm\ge 0$ 
regulate the intensity of the random perturbations
and the Langevin thermostats. 

Finally, we assume that the forcing $\cF_n(t)$ is $\theta_n$-periodic, with the
period $\theta_n=n^{b}\theta$, and the amplitude $n^{-a}$, i.e.
\begin{equation}
\label{Fnt}
 \cF_n(t)= \;\frac{1}{n^{a}} \cF\left(\frac{t}{ n^{b}\theta}\right) + \bar F
\end{equation}
 where $ \cF(t)$ is a smooth $1$-periodic function
such that
\begin{equation}
  \label{eq:2}
  \int_0^1  \cF(t) \dd t = 0, \qquad  \int_0^1  \cF(t)^2 \dd t > 0.
\end{equation}
The constant part of the forcing $\bar F$ does not influence the macroscopic
behavior of   energy in the pinned case, but it is important in the unpinned case
where tension of the chain is a relevant parameter, as it can be seen
in \ref{sec:unpinn-dynam-omeg} below.



In order to ensure stability of the system in the limit $n\to\infty$
we need to assume that the parameters $a,b$ satisfy
\begin{equation}
  \label{eq:3a}
 a\ge 0,\qquad b\ge 0,\qquad   b+a = \frac{1}{2}.
\end{equation}
{The choice $b=0$ corresponds to a fixed period $\theta$ independent of $n$,
  i.e.~the periodic forcing acts on a microscopic time, then we need to set $a=\frac 12$
  in order to have an average work done of order $\frac 1n$ (see \eqref{eq:5c} for
  the definition). In fact the maximum of energy current the system can hold without exploding should be of order $\frac 1n$, and the work done has to be of the
  same order. With the
  choice $0<b<\frac 12$, the periodic forcing is acting on mesoscopic time scales
  and in order to keep the average work done of order $\frac 1n$ we need to choose
  $a = \frac 12 - b$. }

\section{The pinned dynamics: $\omega_0 > 0$ }
\label{sec:pinned-dynamics}

In the presence of the pinning force, $\omega_0> 0$, the
system is not translation invariant and the only conserved quantity in the
bulk is   the energy.

The microscopic energy currents are given by
\begin{equation}
\label{eq:current}
  \frac{d}{dt} \mathcal E_x(t)  = j_{x-1,x}(t) - j_{x,x+1}(t) + \delta_{x,[n\bar u]} \cF_n(t)    p_x(t) ,
\end{equation} 
with 
\begin{equation}
j_{x,x+1}(t):=- p_x(t) (q_{x+1}(t) - q_x(t)) , \qquad \mbox{if }\quad x \in
\{0,...,n-1\}
\label{eq:4}
\end{equation}
and at the boundaries 
  \begin{equation} \label{eq:current-bound}
     j_{-1,0} (t) := 2 { \gamma_-} \left(T_- - p_0^2(t) \right),
      \qquad
  j_{n,n+1} (t):=  -  2 { \gamma_+} \left(T_+ - p_n^2(t) \right).
\end{equation}
The work done up to time $t$ by the periodic force is given by
\begin{equation}
  \label{eq:5c}
  W_n(t) = \int_0^t \cF_n(s)    p_{[nu]}(s) \dd s, 
\end{equation}
where we adopt the usual sign convention that positive work
means energy going into the system.

Consider an initial configuration given by $({\qv,\pv})$, and denote by
$\bbE=\bbE_{\qv,\pv}$ the expectation of the process with this initial configuration.
Thanks to the assumption \eqref{eq:3a} we expect that,  for large $n$, the average
work per unit time is of order $1/n$. In fact, the limit can be
  computed explicitly and for diffusive times $n^2t$ equals:
\begin{equation}
  \label{eq:limW}
  \lim_{n\to\infty} \frac 1n \bbE_{\qv,\pv} \left(W_n(n^2 t)\right)  =
  t \mathbb W,\quad t>0,
\end{equation}
where $\mathbb W$ is independent of ${\qv,\pv}$,
$n$ and $t$. More precisely it is given by
(cf.~\cite[Theorem 2.1 and Remark 2.3]{klo22-2}):
\begin{equation}
  \label{eq:7Q}
 \mathbb W = \left(\frac {2\pi}{\theta}\right)^2   \sum_{\ell\in\bbZ} \ell^2{\cal Q}(\ell),
\end{equation}
where $ {\cal Q}(\ell)$ is explicit: if $b=0$ (and $a=\frac12$) then
\begin{equation} 
\label{021205-21f1}
\begin{aligned}
   {\cal Q}(\ell)= 4\gamma|\hat    \cF(\ell)|^2
  \int_0^1 & \cos^2\left(\frac{\pi z}{2}\right)
  \\ & \times \left\{\left[4\sin^2\left(\frac{\pi z}{2}\right)
      +\om_0^2 -\left(\frac{2\pi\ell}{\theta}\right)^2\right]^2
    +\left(\frac{4 \gamma \pi \ell}{\theta}\right)^2 \right\}^{-1}\dd z
\end{aligned}
\end{equation}
while if $b>0$ then
\begin{equation} 
\label{021205-21f2}
\begin{aligned}
   {\cal Q}(\ell)= 4\gamma|\hat \cF(\ell)|^2
   \int_0^1 \cos^2\left(\frac{\pi z}{2}\right)
   \left[4\sin^2\left(\frac{\pi z}{2}\right)
       +\om_0^2  \right]^{-2}\dd z.
\end{aligned} 
\end{equation}
Note that the latter case 
 corresponds to \eqref{021205-21f1} with  $\lim_{n\to+\infty}\theta_n = +\infty$.
Here
\begin{equation}
\label{cF}
\hat    \cF(\ell)=\int_0^1 e^{-2\pi i\ell t }\cF(t)\dd t,\quad \ell\in\bbZ.
\end{equation}
Note that by \eqref{eq:2} we have $\hat   \cF(0)=0$.
We moreover assume that
\begin{equation}
  \label{eq:18}
  \sum_\ell |\hat     \cF(\ell)| < \infty.
\end{equation}
Notice that $\mathbb W>0$ if
$\sum_{\ell\neq 0} |\hat     \cF(\ell)|> 0$  and it does not depend on
$T_\pm$ nor on $\bar u$.

In \cite{klo22-2} {we have} studied the macroscopic evolution of the temperature profile
in the diffusive space-time scaling in the case $\gamma_+ = 0$
and the periodic force acting on the last particle (i.e.~$\bar u = 1$).
We have assumed that the initial configuration of the particles
is random with a distribution satisfying an entropy bound. More precisely,
define the Gibbs measure
\begin{equation} \label{eq:nuT}
  \begin{split}
    &\nu_{T_-} (\dd{\bf q},\dd{\bf p}) : =
    \frac{1}{Z}\prod_{x=0}^n \exp\left\{-\frac{\mc E_x({\bf
      q},{\bf p})}{T_-} \right\} \dd{\bf q}\dd{\bf p},
\end{split}
\end{equation}
where $Z$ is the normalizing constant.
Let $\mu_n(t)$ be the probability law  of $(\qv(n^2t), \pv(n^2t))$. We
  suppose 
that the initial distribution $\mu_n(0)$ has a density $f_n(0)$ with respect
to $\nu_{T_-}$ that belongs to  $C^2(\bbR^{2(n+1)})$ -- the space of
functions with two continuous derivatives.    By the
  standard regularity theory for SDEs   then $\mu_n(t)$ possesses a
  $C^2$ regular density $f_n(t, \qv, \pv)$ with respect to $\nu_{T_-}$.
We then denote by 
 \begin{equation}
  \label{eq:7a}
\mathbf{H}_{n,T_-}(t) :=  \int_{\Om_n} f_n(t,\qv,\pv)\log f_n(t,\qv,\pv)  \nu_{T_-}(\dd\qv,\dd\pv)
\end{equation}
 the relative entropy of $\mu_n(t)$ w.r.t.~$\nu_{T_-}$.
We assume that there exists a constant $C>0$ such that the relative entropy satisfies 
\begin{equation}
\label{eq:ass0}
    {\mathbf{H}}_{n,T_-}(0) \le C n\quad \mbox{for all $n\ge 1$}.
  \end{equation}
 Furthermore  we suppose that there exists a continuous function
  $T_0:[0,1]\to(0,+\infty)$ such that
\begin{equation}
  \label{eq:6}
  \lim_{n\to\infty} \frac 1{n+1} \sum_x \varphi\left(\frac x{{n+1}} \right)
    \bbE\big(\mathcal E_x(0)\big)
    = \int_0^1 \varphi (u) T_0(u) \dd u,
  \end{equation}
for any   $\varphi\in C[0,1]$ -- the space of continuous
functions on $[0,1]$.
Here and in the following we denote
$\mathcal E_x(t) = \mc E_x({\bf q}(t),{\bf p}(t))$.
Following the same argument as in \cite{klo22-2}, and assuming that
  $\gamma_+>0, \gamma_->0$ (both heat bath are present), we find that  
\begin{equation}
    \label{eq:3bis}
    \lim_{n\to\infty} \frac 1{n+1}
    \sum_x \varphi\left(\frac x{n+1} \right) \bbE
    \left({\cal E}_x(n^2 t)\right) 
    = \int_0^1 \varphi (u) T(t,u) \dd u,
  \end{equation}
  where $T(t,u)$ is {the solution of the heat equation}
  \begin{equation}
    \label{eq:5a}
    \begin{split}
      \partial_t T  = D_\gamma
      \partial_u^2 T, \quad u\in(0,1),
         \end{split}
       \end{equation}
       with $T(0,u)  = T_0(u)$ and with the following boundary conditions:
   \begin{itemize} \item if $\bar u \in (0,1)$
\begin{equation}
    \label{eq:16}
    \begin{split}
      &T(t,0) = T_-, \quad T(t,1) = T_+,\\
     & \partial_u T (t,\bar u^-) - \partial_u T (t,\bar u^+) = \frac{ \mathbb W}{D_\gamma},
    \end{split}
  \end{equation}
 \item if $\bar u=0$ or $1$, then
  the force does not influence
  the hydrodynamic limit,
  since all the energy generated by the work flows into the
  corresponding heat bath, and the boundary conditions are only given by
  \[T(t,0) = T_-,\; T(t,1) = T_+.\] \end{itemize}
Moreover, in the case one thermostat is absent, say $\gamma_+ = 0$, the boundary conditions
  become the following: \begin{itemize}
  \item if $\bar u \in(0,1)$
  \begin{equation}
    \label{eq:16b}
    \begin{split}
    &  T(t,0) = T_-, \quad \partial_u T (t,1) = 0,\\
&      \partial_u T (t,\bar u^-) - \partial_u T (t,\bar u^+) = \frac{ \mathbb W}{D_\gamma},
    \end{split}
  \end{equation}
  \item if, for instance, $\bar u = 1$ then we have
  \begin{equation}
    \label{eq:5}
    \begin{split}
      &T(t,0) = T_-, \quad
      \partial_u T (t,1) = \frac{\mathbb W}{D_\gamma}.
    \end{split}
  \end{equation}\end{itemize}
  This last case \eqref{eq:5} is proven in  \cite{klo22-2}, while the  boundary
  conditions \eqref{eq:16} and \eqref{eq:16b} can be
  proved by a very similar argument.

  The diffusion coefficient $D$ {is not influenced by}  the boundary conditions
  and it is given {in all cases} explicitly  by the formula
   \begin{equation}
     D_\gamma = \frac {1}{4\gamma} \omega_0^2 \Big(G_{\om_0}(0)+ G_{\om_0}(1)\Big)
     = \frac {1}{4\gamma} \frac{2}{2+\om_0^2+\om_0\sqrt{\om_0^2+4}}.
    \label{eq:13}
  \end{equation}
  Here $G_{\om_0}(x) = \left(\omega_0^2 - \Delta\right)^{-1}(x)$, where $\Delta$
  is the  standard discrete Laplacian on $\mathbb Z$.
  The thermal diffusion coefficient $D_\gamma$ can  also be expressed by a
  different formula, that arises from the
  kinetic limit, related to this model \cite{bos09}. Namely,
  \begin{equation}
    \label{eq:8}
    D_\gamma = \frac {1}{4\gamma} {2\pi^2} \int_0^1 \left[\omega'(k)\right]^2 dk,
  \end{equation}
  where $\omega(k) = \sqrt{\omega_0^2 + 4 \sin^2(\pi k)}$ is the dispersion relation
  of the nearest neighbor pinned harmonic chain. The expression
  \eqref{eq:8} gives a general formula for the thermal diffusion for more general
  harmonic chains characterized by the dispersion relation $\omega(k)$.

  In  \cite{klo22-2} it is also proven an equipartition law for  both the
  kinetic and potential energies. It
  implies in particular that
  the limit for the temperature profile equals twice the limit of  the average of
  the kinetic energy, i.e.
\begin{equation}
  \label{eq:6k}
  \lim_{n\to\infty} \frac 1{n+1}
    \sum_x \varphi\left(\frac x{n+1} \right) \bbE
    \left(p^2_x(n^2 t)\right) 
    = \int_0^1 \varphi (u) T(t,u) \dd u,
  \end{equation}

  \subsection{Clausius inequality}
\label{sec:clausius-inequality}




From the evolution of the relative entropy, if $\gamma_+ = 0$ we get the following inequality
{
\begin{equation}
  \label{eq:9}
  \begin{split}
    \frac 1n \left(\mathbf{H}_{n,T_+}(t) - \mathbf{H}_{n,T_+}(0)\right)
    \le \frac 1{nT_-} \mathbb E\left(W_n(n^2 t)\right)
    .
\end{split}
\end{equation}
By  Fourier's law  (see \eqref{eq:5a})
the macroscopic current at the right endpoint
equals
$-(D/4\ga)\int_0^t\partial_uT(s,1)\dd s$, 
and we obtain the
following inequality in the macroscopic limit:
\begin{align}
  \label{eq:clausius}
  &
 \lim_{n\to+\infty} \frac 1n \left(\mathbf{H}_{n,T_+}(t) -
    \mathbf{H}_{n,T_+}(0)\right) \le \frac{t \mathbb W}{T_-}.
\end{align}
}

\subsection{Stationary periodic state}

{ We define
a \emph{periodic stationary probability measure}
$\{\mu_t^P, t\in[0,+\infty)\}$ for the dynamics of the chain as
a solution of the forward equation $\partial_t \mu_t^P = \mathcal G_t^* \mu_t^P$
such that $\mu_{t + \theta_n}^P=\mu_{t}^P$, where $\mathcal G_t^*$ is the adjoint of the
generator $\mathcal G_t$ of the dynamics.
This condition is equivalent with 
\begin{equation}
  \label{eq:23}
  \int_0^{\theta_n} \dd s \int_{\bbR^{2(1+n)}} \mathcal G_s F(\rv, \pv) \mu_s^P(\dd\qv,\dd\pv)
    = 0,
\end{equation}
for any smooth test function $F $. Using the
contraction principle, in a manner similar to the proof of the
existence and uniqueness of
self-consistent reservoirs for a harmonic crystal (see \cite[Theorem 3.1]{bll}) one can prove that
for a fixed $n\ge1$ there exists a unique $\theta_n$-periodic stationary state $\{\mu_s^P,
s\in[0,+\infty)\}$ for the system \eqref{eq:flip}-\eqref{eq:pbdf}.
 The measures $\mu_s^P$ are absolutely continuous with respect to the Lebesgue
 measure $\dd\qv\dd\pv$ and the density
 $\mu_s^P(\dd\qv,\dd\pv)=f_s^P(\qv,\pv) \dd\qv\dd\pv$ is
 strictly positive. This   has been shown in the case $\ga_+=0$
 and $\bar u=1$ in \cite[Theorem 1.1]{klo22}.

 Suppose that  $\{(\mathbf q(t), \mathbf p(t))\}_{t\ge0}$ is the
solution of \eqref{eq:flip}-\eqref{eq:pbdf}  initially distributed
according to $\mu_0^P$.
Given a measurable function $F:\bbR^{2(n+1)}\to\bbR$ integrable
w.r.t. each measure $\{\mu_s^P,
s\in[0,+\infty)\}$ we denote
\begin{equation}
\label{bar}
\bar F(t):=\bbE_{\mu_0^P}\Big(F\big(\mathbf q(t), \mathbf
p(t)\big)\Big)=\int_{\bbR^{2(n+1)}}F(\qv,\pv)\mu_t^P(\dd\qv,\dd\pv),\quad t\ge0,
\end{equation}
where $\bbE_{\mu_0^P}$ is the expectation    {  corresponding to
  the    initial data distributed according to $\mu_0^P$.}
The function $\bar F(t)$ is $\theta_n$-periodic.
We denote its time average by
\begin{equation}
\label{ll}
\lang F\rang:= {\frac{1}{\theta_n}\int_0^{\theta_n}\bar F(t)\dd t}.
\end{equation}}

\subsection{The macroscopic stationary state}
\label{sec:stat-peri-state}

{
In the general case the stationary temperature profile, {corresponding
to \eqref{eq:5a} and \eqref{eq:16}}, is given by
\begin{equation}
  \label{eq:17}
  \begin{split}
    T_{ss}(u) = &\left[ T_- + \left(\frac{\mathbb W}{D_\gamma} (1-\bar u)
      + T_+ - T_-\right) u\right] 1_{u\le \bar u}
  \\
  &
  + \left[ T_+
    + \left(\frac{ \mathbb W}{D_\gamma} \bar u - T_+ + T_-\right) (1-u)\right] 1_{u > \bar u}.
\end{split}
\end{equation}
If the right heat bath is absent, i.e.~$\gamma_+ = 0$ {(then the
boundary condition \eqref{eq:16b} holds)}, then
\begin{equation}
  \label{eq:20}
   T_{ss}(u) = \left[ T_- + \frac{\mathbb  W}{D_\gamma}  u\right] 1_{u\le \bar u}
  + \left[ T_- +  \frac{\mathbb  W}{D_\gamma} \bar u \right] 1_{u > \bar u}.
\end{equation}
Finally, if $\gamma_+= 0$ and $\bar u =1$ {corresponding to the
boundary condition \eqref{eq:5}}, then it has been proved in
\cite[Theorem 3.3]{klo22} that for any $\varphi\in C[0,1]$:
\begin{equation}
    \label{eq:3}
   \begin{aligned} \lim_{n\to\infty} \frac 1{n+1} \sum_{x=0}^n \varphi\left(\frac x{n+1} \right) \lang p^2_x\rang
     & =  \lim_{n\to\infty} \frac 1{n+1} \sum_{x=0}^n \varphi\left(\frac x{n+1} \right)
    \lang {\cal E}_x\rang \\ & 
    = \int_0^1 \varphi (u) T_{ss}(u) \dd u, \end{aligned}
  \end{equation}
 with 
the stationary profile  given by 
\begin{equation}
  \label{eq:20b}
  T_{ss}(u) = T_- + \frac{\mathbb  W}{D_\gamma}  u.
\end{equation}
{Notice that when $\gamma \to 0$ we have that $T_{ss}(u) \to T_-$ in
  \eqref{eq:20} and \eqref{eq:20b}, while $T_{ss}(u) \to (T_+-T_-)u + T_-$ in
  \eqref{eq:17}. This does not correspond to the stationary situation
  with $\gamma = 0$, cf. \cite{RLL67} for the case in absence of periodic forcing.}

Furthermore, in  \cite[Theorem 9.1]{klo22}, we prove that in the case when the period of the force is of a fixed
  microscopic size (i.e. $b=0$ and $a=-1/2$) the fluctuations of the kinetic energy
functional vanish, i.e.~there exists a constant $C>0$ such that
\begin{equation}
    \label{eq:91cc}
  {    \sum_{x=0}^n \int_0^\theta \left(\bar{p_x^2}(t) - \lang
      p_x^2\rang \right)^2\dd t    
\le \frac{C}{n^2}, \quad n=1,2,\dots}
  \end{equation}}

\section{The unpinned dynamics: $\omega_0 = 0$ }
\label{sec:unpinn-dynam-omeg}

When the system is unpinned, i.e. $\omega_0 = 0$,
it is translational invariant and one should consider only
the relative distance between the particles. We introduce the variables
\begin{equation}
  \label{eq:10}
  r_x:= q_x - q_{x-1}, \qquad x=1,\dots, n,
\end{equation}
sometimes referred to as  the {\em volume stretch}.
{In this situation there are two conserved quantities in the bulk: the energy
  $\E_x = \frac 12(p_x^2 + r_x^2)$ and the volume $r_x$. The hydrodynamic limit
  for this unpinned dynamics with Langevin heat bath at both endpoints
  has been studied in \cite{kos3}.}
In \cite{klos22}  we consider the situation
when the force is applied at the right endpoint of the chain and the
only heat bath is located at its left endpoint. The  microscopic dynamics of
the process $\{(\mathbf r(t), \mathbf p(t))\}_{t\ge0}$
describing the total chain is  given  by
\begin{equation} 
\label{eq:flip1}
\begin{aligned}
  \dot{   r}_x(t) &=  p_x(t) - p_{x-1}(t) \\
  \dd   p_x(t) &=  (r_{x+1}(t)-r_x(t)) \dd t-   2
  p_x(t^-) \dd N_x(\gamma  t) +\delta_{x,n}{\cal F}_n(t)\dd t,  \end{aligned} \end{equation} for $x=1,\ldots,n
$
and at the left boundary 
\begin{align}
     \dd   p_0(t) =    r_1  \dd   t
                     -
                    2  \gamma_- p_0(t) \dd t
                    +\sqrt{4  \gamma_- T_-} \dd w_-(t)
                    \vphantom{\Big(} \label{eq:pbdf1}.
\end{align}
Here the force is given by \eqref{Fnt}. We use
  the convention $r_0=r_{n+1}:=0$. 
  One can immmediately see, from the first equation of \eqref{eq:flip1},
  that $r_x$ is a second (locally) conserved quantity besides the energy.

The energy   currents are again given by 
\eqref{eq:4} and \eqref{eq:current-bound}. The work performed by the
force is again given by \eqref{eq:5c}.
{We have
\begin{equation}
  \label{eq:6up}
  \lim_{n\to\infty} \frac 1n \bbE_{\mathbf{r},\pv} \left(W_n(n^2 t)\right)  = t \mathbb W
\end{equation}
where $\mathbb W$ is independent of $t$ and $({\mathbf{r},\pv})$  -- the initial
configuration of stretches and momenta.
In the case $a=1/2$ and $b=0$ it is given by (see \cite{klos22} in preparation):
\begin{equation}
  \label{eq:7}
  \mathbb W =  \mathbb W_{\rm mech}+\mathbb W_{\rm Q},
\end{equation}
where 
\begin{align*}
  \mathbb W_{\rm mech}:= \frac {\bar F^2}{2\gamma} ,\qquad
    \mathbb W_{\rm Q}:= \sum_{\ell\in\bbZ} \left(\frac {2\pi \ell}{\theta}\right)^2{\cal Q}(\ell),
\end{align*}
correspond to the mechanical and thermal parts of the work performed
on the system.  }
Here ${\cal Q}(\ell)$ is given by 
\eqref{021205-21f1}, setting $\om_0=0$. In the case  $b>0$, we have
$a+b/4=1/2$ and $\mathbb W$ is given by \eqref{eq:7} with the same
formula for $\mathbb W_{\rm mech}$ and 
$$
 \mathbb W_{\rm Q}:=2\sum_{\ell\in\bbZ} \big(\frac{\pi 
     | \ell|}{\ga\theta} \big)^{1/2}|\hat {\cal
    F}(\ell)|^2 .
$$
To formulate
the hydrodynamic limit we assume, besides \eqref{eq:ass0} and \eqref{eq:6}, that
   for any test function $\varphi\in C[0,1]$ we have
\begin{equation}
\label{weak-r0-p0} \begin{aligned}
&\lim_{n\to+\infty}\frac{1}{n+1}\sum_{x=0}^n \bbE\big(p_x^2
(0)\big)\varphi\left(\frac{x}{n+1}\right)=\int_{0}^1T_0(u)\varphi(u)\dd
u,\\
&\lim_{n\to+\infty}\frac{1}{n+1}\sum_{x=0}^n 
\bbE\big(r_x (0)\big)\varphi\left(\frac{x}{n+1}\right)=\int_{0}^1r_0(u)\varphi(u)\dd u.
\end{aligned} \end{equation}
In addition, {if} we assume that $a=1/2$ and $b=0$,
{then}
$$
\Big(\bbE \big(p^2_x(n^2t)\big),\bbE \big(r_x(n^2t)\big)\Big)_{x=0,\ldots,n}
$$ converge 
weakly, cf.~\eqref{eq:3},  to $(T(t,u),r(t,u))$, the unique solution of
the following system
\begin{align}
  &\partial_t T(t,u) = \frac{1}{4\gamma}\partial_{uu} T(t,u)
    + \frac{1}{2\gamma} (\partial_u r(t,u))^2 \notag\\
       &
         \partial_t
         r(t,u)=\frac{1}{2\gamma}\partial_{uu}r(t,u), \qquad  (t,u)\in \R_+\times (0,1),
         \label{eq:linear} 
      \end{align}
 with the  boundary and initial conditions:
\begin{equation}
  \label{eq:bc0}
  \begin{split}
   & r(t,0) = 0  , \qquad  r(t,1) = \bar F,\\
   & T (t,0) = T_-, \quad \partial_u T (1) = 4\gamma \mathbb{ W}_Q\\
&T(0,u)= T_0(u),\quad   r(0,u)=r_0(u).
  \end{split}
\end{equation}  
This result {will be} shown in \cite{klos22}.

{ Notice that the thermal energy, i.e.~the temperature $T(t,u)$,
  is not a conserved quantity in the bulk. This is given instead by the total energy
$\E(t,u) = T(t,u) + \frac{1}{2}r(t,u)^2$ that satisfies the equation
\begin{equation}
  \label{eq:29}
  \partial_t \E(t,u) = \frac{1}{4\gamma}\partial_{uu} \left(\E(t,u)
    + \frac{1}{2}  r(t,u)^2\right)
\end{equation}
that is equivalent to \eqref{eq:linear}. Consequently we understand that the term
$\frac{1}{2\gamma} (\partial_u r(t,u))^2$ is the rate of transfer of mechanical energy
to thermal energy in the bulk. Of course we also have that $\bbE \big(\E_x(n^2t)\big)$
converges weakly to $\E(t,u)$.
}

In the case the forcing is done on a point $[n\bar u]$ in the bulk of the system,
and a heat bath is present on the right hand side ($\gamma_+>0$),
then the boundary conditions we expect are the following:
\begin{equation}
  \label{eq:bc1}
  \begin{split}
    & r(t,0) = 0  , \quad \partial_u r(t,\bar u^+) = \partial_u r(t,\bar u^-),
   \\ & r(t,\bar u^-) -   r(t,\bar u^+)= \bar F,\quad r(t,1) = 0,\\
   & T (t,0) = T_-, \quad \partial_u T (\bar u^-) - \partial_u T (\bar u^+)
   = 4\gamma \mathbb{ W}_Q, \qquad T (t,1) = T_+.\\
  \end{split}
\end{equation}


{As in the pinned case, the macroscopic stationary temperature
  profiles can be computed. In the case of the stationary state
  corresponding to \eqref{eq:linear}   the elongation stationary profile is given by 
  \begin{equation}
    \label{extra1}
    r_{ss}(u)=\overline{F}u, \qquad u \in [0,1]\end{equation}
and the temperature stationary profile is given by
 \begin{equation}
    \label{extra2} T_{ss}(u)= \overline{F}^2 u(1-u)+(\overline{F}^2+4\gamma \mathbb{W}_Q)u
    +T_-, \qquad u \in [0,1].
    \end{equation}
    Note that, contrary to the pinned case, {the temperature profile}
    is not linear (see \eqref{eq:20b}) but parabolic.
}

In the case the forcing is in the bulk and the thermostats are present
at both
endpoints, the stationary solution with boundary conditions
\eqref{eq:bc1} is given by
\begin{equation}
  \label{eq:28}
  \begin{split}
    r_{ss} (u) = &\bar F \left( u - 1_{u\ge \bar u}\right),\\
    T_{ss} (u) =  &\left[ T_- + \left({4\gamma \mathbb W_Q} (1-\bar u)
      + T_+ - T_-\right) u\right] 1_{u\le \bar u}
  \\
  &
  + \left[ T_+ +
    \left({4\gamma \mathbb W_Q} \bar u - T_+ + T_-\right) (1-u)\right] 1_{u > \bar u}
  + \overline{F}^2 u(1-u).
  \end{split}
\end{equation}

\label{sec:stat-peri-state-1}

\section{Higher dimension}
\label{sec:d-dimension}

We can consider the discrete lattice
$$
\Xi_{d,n} = \{ \bx = \{x_1,\dots, x_d\}, x_j= 0,\dots,n,
x_j = x_{j+n} \ \text{if}\ j\neq 1\},
$$
and the configuration of positions and momenta are described by
$$
 (\mathbf q, \mathbf p) = (q_\bx, p_\bx)
  \in \R^{\Xi_{d,n}}\times\R^{\Xi_{d,n}}. 
  $$
  The  microscopic dynamics of
the process $\{(\mathbf q(t), \mathbf p(t))\}_{t\ge0}$
describing the total chain is now given in the bulk by
\begin{equation} 
\label{eq:flipd}
\begin{aligned}
  \dot   q_\bx(t) &= p_\bx(t) ,
  \qquad  \bx\in \Xi_{d,n},\\
  \dd   p_\bx(t) &=  \left(\Delta_N q_\bx-\om_0^2 q_\bx\right) \dd t-   2
  p_\bx(t-) \dd N_\bx(\gamma t) + \cF_{n,\bx}(t)
                 \dd t  ,
  \end{aligned} \end{equation}
for $x_1\in \{1, \dots, n-1\},$ and at the boundaries by 
\begin{align}
     \dd   p_\bx(t) &=   \; \Big(\Delta_N q_\bx - \om_0^2 q_\bx(t) \Big) \dd   t
                     - 2  \gamma_- p_\bx(t) \dd t
                    +\sqrt{4  \gamma_- T_-} \dd \tilde w_\bx(t), x_1= 0
             \notag      \\
  \dd   p_\bx(t) &=  \; \Big(\Delta_N q_\bx -\om_0^2 q_\bx(t) \Big)  \dd   t
                 -  2  \gamma_+ p_\bx(t) \dd t +\sqrt{4  \gamma_+ T_+}\dd \tilde w_{\bx}(t),
                      x_1 = n. 
 \label{eq:pbdfd}
\end{align}
Here $\Delta_N$ is the discrete laplacian on $\Xi_{d,n}$
with Neumann boundary conditions on the direction $1$ and periodic on the others.
Processes $\{N_\bx(t), \bx\in\Xi_{d,n}\}$
are independent Poisson of intensity $1$, while
$\{\tilde w_\bx(t)\}$ are independent standard Wiener processes,
independent of the Poisson processes.


For the pinned model ($\om_0 \neq 0$) in the absense of forcing
($\cF_{n,\bx}(t) = 0$ for all $\bx\in \Xi_{d,n}$),
the macroscopic heat equation is given by
 \begin{equation}
    \label{eq:5d}
    \begin{split}
      &\partial_t T  = D_\gamma 
      \Delta_u T, \quad u\in(0,1)\times \TT^{d-1},\\
      &T(t,u) = \begin{cases} T_- & \; u_1 = 0,\\ T_+ &\; u_1 = 1,\end{cases}
    \end{split}
  \end{equation}
  where the diffusion coefficient depends on the dimension and is given by
  \begin{equation}
    \label{eq:19}
    D_\gamma = \frac{1}{4\gamma}\frac 1{2 \pi^2}
    \int_{\TT^d} \left|\nabla \omega(\mathbf k)\right|^2 d \mathbf k
  \end{equation}
  where $\omega(\mathbf k) = \sqrt{\omega_0^2 + 4\sum_{j=1}^d \sin^2(\pi k_j)}$
  is the dispersion relation of the harmonic lattice.
  This coincides with the diffusion coefficient appearing in the self-consistent model,
  computed in section 7 of \cite{bll}.

  If a driving force is acting in the same way on the right side of the system, i.e.~$\mathcal F_{n,\bx}(t) = \mathcal F_{n}(t) \delta_{x_1, n-1}$, with $\mathcal F_{n}(t)$
  satisfying analogous conditions as in dimension 1, and in absence
  of thermostat on the right, i.e.~$\gamma_+ = 0$, then we have
  the following Neumann type boundary conditions on the right:
  \begin{equation}
    \label{eq:24}
    \nabla T(t,1) = \frac{\mathbb W}{D_\gamma}\;
    \begin{pmatrix}
      1\\ 0\\ \vdots\\ 0
    \end{pmatrix}
  \end{equation}
  where $\mathbb W$ is the average work done by the periodic forcing,
  defined as the limit
  \begin{equation}
    \label{eq:25}
    \begin{split}
      \mathbb W=
      \frac 1t \lim_{n\to\infty} \frac 1{n^{d-1}} \sum_{\bx}\delta_{x_1, n-1}
      \mathbb E \left( \frac 1n \int_0^{n^2 t} \mathcal F_n(s) p_\bx(s) ds\right).
    \end{split}
  \end{equation}
  More generally, given a smooth $d-1$ dimensional  surface
  $\Gamma \subset (0,1)\times \TT^{d-1}$, defining by $\Gamma_n$ the lattice points
  $\bx$ such that the distance of $\bx/n$ from $\Gamma$ is less that $\frac 1{2n}$, 
  if
  $\mathcal F_{n,\bx}(t) = \mathcal F_{n}(t) \delta_{\bx, \Gamma_n}$,
  then we expect the boundary conditions
  \begin{equation}
    \label{eq:26}
    \nabla T(t,u) = \frac{ \mathbb W}{D_\gamma} \mathbf n_\Gamma(u),
    \qquad u\in \Gamma,
  \end{equation}
  where $\mathbf n_\Gamma(y)$ is the unit vector normal to $\Gamma$ in the point $u$,
  and $\mathbb W$ is a suitable modification of formula \eqref{eq:25}.
  For a more general inhomogeneous periodic forcing
  different macroscopic boundary conditions
  are expected, and it is a subject of further investigation.  The argument used in the one dimensional case in \cite{klo22,klo22-2} cannot be extended directly to prove
  \eqref{eq:5d}, \eqref{eq:24} or \eqref{eq:26}.
  This will be considered in a future work (in preparation).

\section{Anharmonic chains}
\label{sec:anharmonic-chains}

Harmonic chains allow many explicit calculations, in particular we can solve
first and second moment equations autonomously,
without any need to analyze higher moments.
The situation is much more difficult for anharmonic chains, even in presence of the random flip of the velocities sign.
Consider the Hamiltonian
\begin{equation}
  \label{eq:11}
  \mathcal{H}_n (\mathbf q, \mathbf p):=
\sum_{x=0}^n \left(\frac{p_x^2}{2} + V(q_x-q_{x-1}) + U(q_x)\right),
\end{equation}
where we set $q_{-1}= 0$.
Then consider the stochastic dynamics
\begin{equation} 
\label{eq:flipnh}
\begin{aligned}
  \dot   q_x(t) &= p_x(t) ,
  \qquad \qquad \qquad  \qquad \qquad x\in \{0, \dots, n\},\\
  \dd   p_x(t) &=  \partial_{q_x} \mathcal{H}_n \dd t-   2
  p_x(t-) \dd N_x(\gamma t),
  \quad x\in \{1, \dots, n-1\},
  \end{aligned} \end{equation}
and at the boundaries by 
\begin{align}
  \dd   p_0(t) &=   \; \partial_{q_0} \mathcal{H}_n \dd   t
                 -2  p_0(t-) \dd N_0(\gamma t)
                     - 2  \gamma_- p_0(t) \dd t
                    +\sqrt{4  \gamma_- T_-} \dd \tilde w_-(t)
                    \vphantom{\Big(}\notag  \\
  \dd   p_n(t) &=  \; \partial_{q_n} \mathcal{H}_n \dd   t  +\cF_n(t)
                 \dd t  -2  p_n(t-) \dd N_n(\gamma t)
                 - 2  \gamma_+ p_n(t) \dd t \notag \\ & \quad +\sqrt{4  \gamma_+ T_+}\dd \tilde w_+(t).
                     \vphantom{\Big(}  
\label{eq:pbdfnh}
\end{align}
The energy currents in the bulk are given by
\begin{equation}
  j_{x,x+1}(t):=- p_x(t) V'(q_{x+1}(t) - q_x(t)) ,
  \qquad \mbox{if }\quad x \in \{0,...,n-1\}.
\label{eq:44}
\end{equation}
The only existing mathematical result is the existence of the thermal diffusivity defined
by the Green-Kubo formula (cf. \cite{bo1}):
\begin{equation}
  \label{eq:12}
  \begin{split}
    D_\gamma(T) = \int_0^\infty \dd t \sum_{x\in\bbZ}
    \mathbb E_{\nu_T}\left( j_{x,x+1}(t)  j_{0,1}(0)\right),
  \end{split}
\end{equation}
where $\mathbb E_{\nu_T}$ is the expectation of the corresponding infinite dynamics
in equilibrium at temperature $T$.

When $\gamma_-$ and $\gamma_+$ are strictly positive we expect the convergence
of the temperature profile as in \eqref{eq:6k} to the solution of
\begin{equation}
  \label{eq:14}
   \begin{split}
      &\partial_t  T  = \partial_u\left(D_\gamma(T) \partial_u T\right), \qquad u\in(0,1),\\
      &T(t,0) = T_-, \quad  T (t,1) = T_+,\\&
      T(0,u)  = T_0(u).
    \end{split}
\end{equation}
When
$\gamma_+=0$, i.e.~if only the periodic forcing is acting on the last particle,
the boundary condition on $u=1$ is given by a non-linear Neumann condition:
\begin{equation}
  \label{eq:15}
  D_\gamma(T(t,1)) (\partial_u T)(t,1) = -J(T(t,1))
\end{equation}
with a boundary current $J(T)$ depending on the local temperature.

A linear response argument gives the following expression of $J$ as a function of
$T$:
\begin{equation}
  \label{eq:27}
  \begin{split}
  - J(T) &= \frac{1}{T} \frac 1\theta \int_0^\theta \dd t
  \int_t^{+\infty}\mathcal F\Big(\frac{t}{\theta}\Big) \mathcal F\Big(\frac{s}{\theta}\Big)  
  \mathbb E^+_{\nu_T} \left(p_0(t) p_0(s)\right)\; \dd s\\
  &=  \frac{1}{T} \int_0^\infty  \left(\frac 1\theta \int_0^\theta
    \mathcal F\Big(\frac{t}{\theta}\Big) \mathcal
    F\Big(\frac{t+s}{\theta}\Big)   \dd t\right)
  \mathbb E^+_{\nu_T} \left(p_0(0) p_0(s)\right)\; \dd s,
\end{split}
\end{equation}
where $\mathbb E^+_{\nu_T}$ denotes the expectation for the semi-infinite process
(i.e. \eqref{eq:flipnh} for $x\in \mathbb Z_+$, without any forcing or heat bath)
in equilibrium at temperature $T$. As for \eqref{eq:12}, the integral involved in \eqref{eq:27}
can be proven convergent by using a similar argument as in \cite{bo1}. 

In the harmonic case \eqref{eq:27} coincides with \eqref{eq:7Q}
with $- J(T)=\mathbb W$.
In the case of quartic pinning potential $U$ and quadratic $V$, \eqref{eq:15}
and \eqref{eq:27} have been confirmed recently by numerical simulations
\cite{shiva}.

Note also that $- J(T)$ is positive since
\begin{equation}
  \label{eq:15bis}\begin{split}
  - J(T)&=\lim_{n\to+\infty}\frac{1}{n T}\int_0^{n\theta} \dd t \int_t^{n\theta}{\cal
    F}\left(\frac{t}{\theta}\right){\cal
    F}\left(\frac{s}{\theta}\right) \bbE_{\nu_{T}}
  \Big[ p_n(t) p_n(s)\Big] \dd s\\
&
=
  \lim_{n\to+\infty}\frac{1}{2nT}\bbE_{\nu_{T}}
\Bigg[\left(\int_0^{n\theta} {\cal
    F}\left(\frac{t}{\theta}\right)p_n(t)\dd t\right)^2\Bigg]  \ge0.
\end{split}
\end{equation}

\bibliographystyle{plain}
\bibliography{bibliography}

\index{bibliography}

\end{document}